\documentclass[%
 aip,
 jmp,%
 amsmath,amssymb,
reprint,%
]{revtex4-1} % for review and submission

\usepackage{graphicx}  % needed for figures
\usepackage{dcolumn}   % needed for some tables
\usepackage{bm}        % for math
\usepackage{amssymb}   % for math
\usepackage{amsmath}
\usepackage{float}
\nocite{*}

\begin{document}

% the following line is for submission, including submission to the arXiv!!
%\hspace{5.2in} \mbox{Fermilab-Pub-04/xxx-E}

\title{Proposed scheme to generate bright entangled photon pairs by application of a quadrupole field to a single quantum dot}
%\input author_list.tex       % D0 authors (remove the first 3 lines
                             % of this file prior to submission, they
                             % contains a time stamp for the authorlist)
                             % (includes institutions and visitors)
\date{\today}
\author{M. Zeeshan}
\email{m5zeeshan@uwaterloo.ca}
\author{N. Sherlekar}
\affiliation{Institute for Quantum Computing and Department of Electrical and Computer Engineering, University of Waterloo, Waterloo,
N2L 3G1, Canada}
\author{A. Ahmadi}
\affiliation{Institute for Quantum Computing and Department of Physics and Astronomy, University of Waterloo, Waterloo,
N2L 3G1, Canada}
\author{R.L. Williams}
\affiliation{National Research Council of Canada, Ottawa, Ontario, Canada, K1A 0R6}
\author{M.E. Reimer}
\affiliation{Institute
for Quantum Computing and Department of Electrical and Computer Engineering, University of Waterloo, Waterloo,
N2L 3G1, Canada}

\begin{abstract}
Entangled photon sources are crucial for quantum optics, quantum sensing and quantum communication. Semiconductor quantum dots generate on-demand entangled photon pairs via the biexciton-exciton cascade. However, the pair of photons are emitted isotropically in all directions, thus limiting the collection efficiency to a fraction of a percent. Moreover, strain and structural asymmetry in quantum dots lift the degeneracy of the intermediate exciton states in the cascade, thus degrading the measured entanglement fidelity. Here, we propose an approach for generating a pair of entangled photons from a semiconductor quantum dot by application of a quadrupole electrostatic potential. We show that the quadrupole electric field corrects for the spatial asymmetry of the excitonic wavefunction for any quantum dot dipole orientation and fully erases the fine-structure splitting without compromising the spatial overlap between electrons and holes. Our approach is compatible with nanophotonic structures such as microcavities and nanowires, thus paving the way towards a deterministic source of entangled photons with high fidelity and collection efficiency.
\end{abstract}

\maketitle
%Semiconductor quantum dots in nanowires have recently emerged as promising candidates for such sources considering their bright and directional emission with Gaussian emission profile for the near-unity fiber coupling efficiency.

Quantum dots generate polarization entangled photons on-demand via the biexciton-exciton cascade \citep{Salter2010b,Moreau_entangled,Muller2014,Zhang2017}.
%- The ground state is occupied by two electron-hole pairs
However, an energy splitting of the intermediate exciton states, known as the fine structure splitting (FSS), introduces a which-path information within the biexciton-exciton cascade and reduces the measured polarization entanglement \citep{Bayer2002,Fognini2017a}. This energy splitting can be caused by an asymmetric quantum dot shape \citep{Krapek2015,Ranber_shape}, piezoelectric field \citep{Seguin_piezo}, stress \citep{Zieliski2013} and inhomogeneous alloying \citep{Ranber_atomic}, which reduces the symmetry of the excitonic confinement potential.

Several quantum dot growth techniques have been developed \citep{Kuroda_growth,Huo2013_growth,PhysRevApplied.8.014013,Keil2017} to minimize the FSS, but only a limited number of quantum dots on the sample will have an FSS close to zero. Therefore, post-growth perturbation techniques are introduced to further reduce the FSS such as application of electric fields \citep{Bennett2010}, strain fields \citep{Wang2012}, magnetic fields \citep{Stevenson_magnet,Stevenson_}, annealing \citep{Langbein_annealing} and by a combination of strain and an electric field \citep{Trotta2012,Zhang2017}. Using strain was shown to be the most versatile approach in addressing the challenge of minimizing the FSS towards zero \citep{Trotta2015,Trotta2016a}. However, such post-growth tuning techniques have not been implemented on quantum dots in photonic nanostructures such as nanowires \citep{Claudon2010,Reimer2012a} or micropillar cavities \citep{Gazzano2013} for enhanced photon collection efficiency with near-unity single mode fiber coupling \citep{Bulgarini2014}.

\begin{figure}[ht]
\centering
\includegraphics{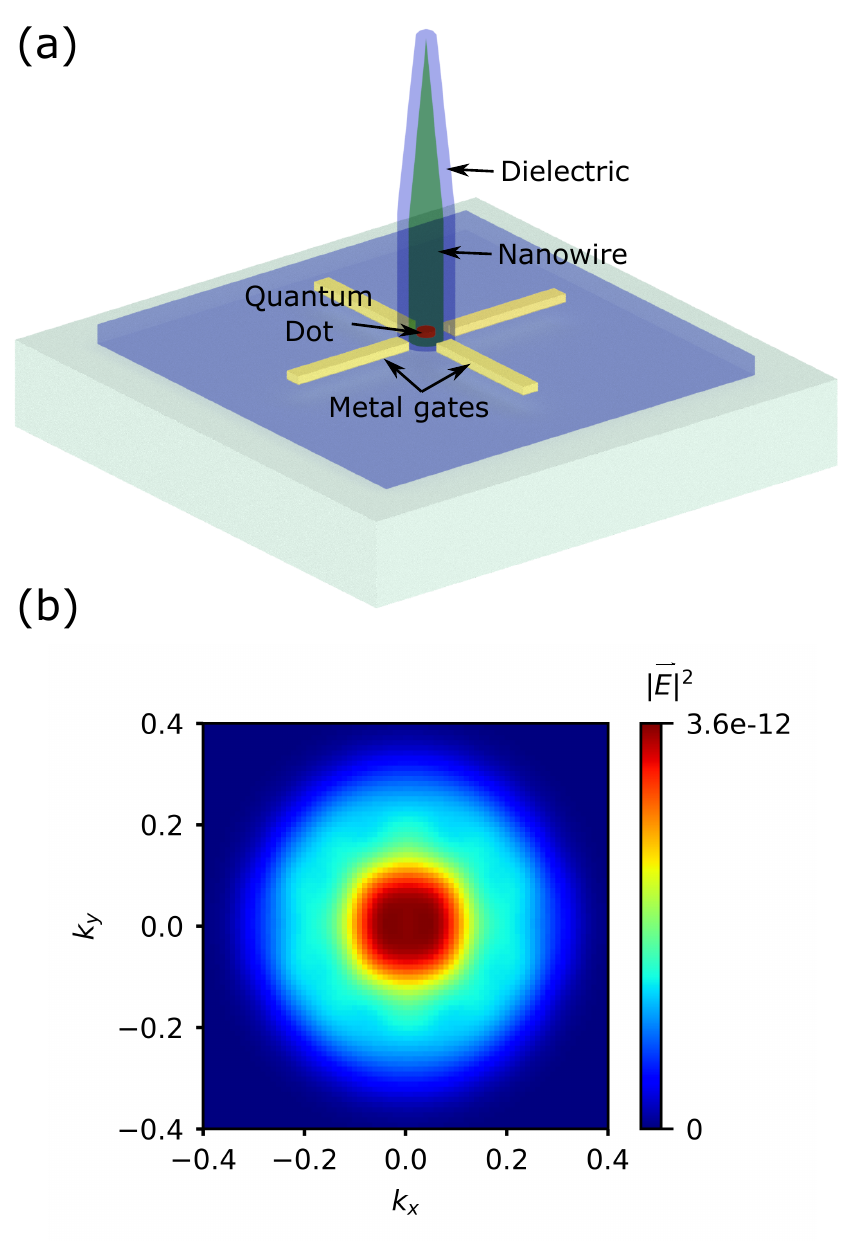}
\caption{\label{fig:device_farfield}Device design. (a)  Schematic view of the proposed device architecture consisting of a single quantum dot in a standing nanowire, dielectric coating and four electrical contacts in-plane of the quantum dot. (b) Calculated far-field emission profile from the device using a finite-difference time-domain method in Lumerical, assuming an in-plane dipole on the nanowire waveguide axis.}
\end{figure}

In this letter, we propose to remove the FSS by applying a quadrupole electric field to a single quantum dot while maintaining a high light extraction efficiency. Fig. 1(a) shows a schematic view of the proposed device, with four gates surrounding a single quantum dot in a tapered nanowire waveguide. The tapered nanowire allows for a high light extraction efficiency, whereas the four gates remove the FSS by applying quadrupole electric field. Using this quadrupole electric field we show that the FSS can be removed for any in-plane quantum dot dipole orientation without compromising the electron-hole (e-h) overlap. Maintaining this strong e-h overlap in a quadrupole field is in stark contrast to previous electric field implementations to remove the FSS. In previous works, application of a lateral electric field resulted in the reduction of e-h overlap and a lower brightness of the quantum dot \citep{Gerardot2007a,Korkusinski2009a,Kowalik2005,
Kowalik2007}.
Finally, we emphasize that the quantum dot dipole orientation should be aligned with the applied lateral field to ensure zero FSS, whereas here we show that this is not necessary for the case of a quadrupole field.

\begin{figure*}[t!]
\centering
\includegraphics{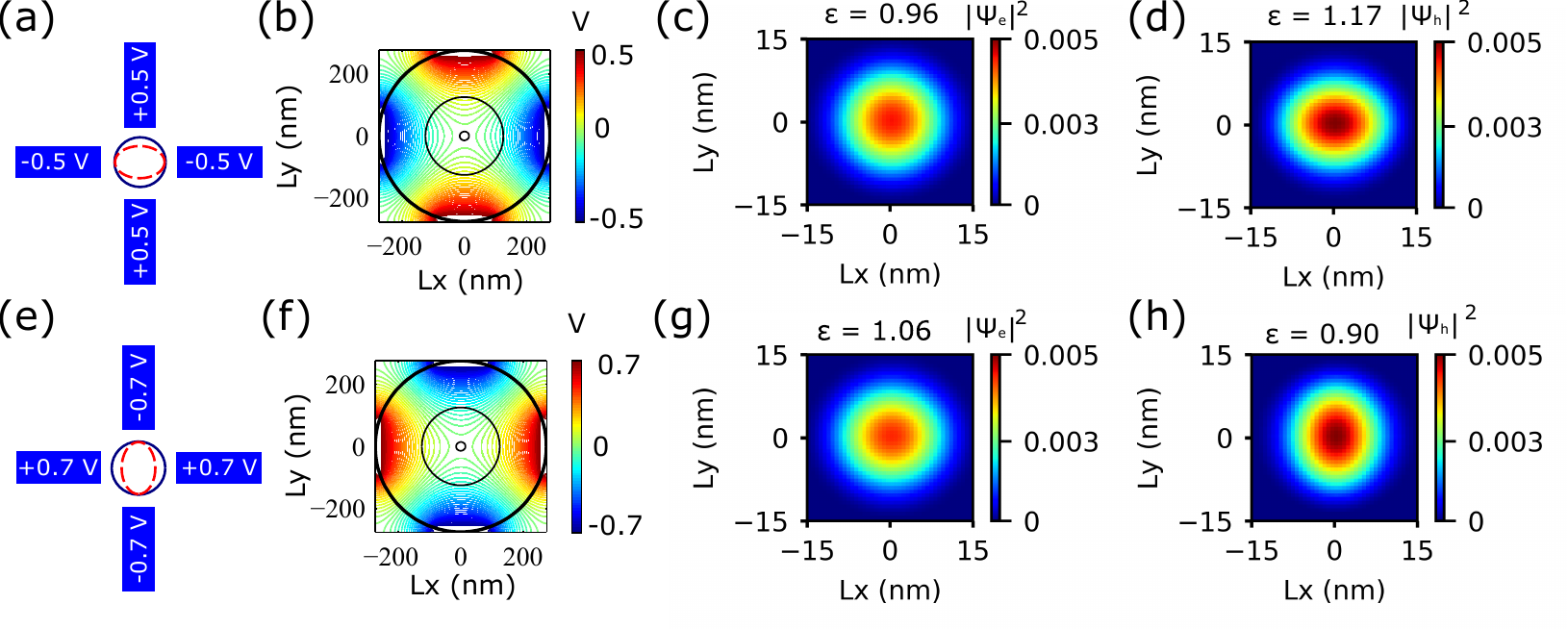}
\caption{\label{fig:wavefunction}Single particle wavefunctions with an applied quadrupole field. (a) Schematic view of applied quadrupole potential. We use the convention that the top and bottom gates are positive at +0.5\,V and the left and right gates are negative at -0.5\,V.  The electron wavefunction is shown schematically in solid blue and hole wavefunction in dashed red. (b) Calculated 2D electric potential for configuration (a). The black circles represent the edges of the dielectric (largest), nanowire (medium) and quantum dot (smallest). (c),(d) Calculated 2D probability density of electron (c) and hole (d) wavefunctions with configuration from (a). (e) Schematic view of applied quadrupole potential with polarities reversed with respect to (a) and magnitude increased to 0.7\,V. (f) Calculated electric potential for configuration from (e). (g),(h) Calculated 2D probability density of electron (g) and hole (h) wavefunction with configuration from (e). Lx and Ly are the dimensions in the x and y direction of the simulated structure, whereas $\epsilon$ is the degree of single particle wavefunction elongation of the electron and hole. }
\end{figure*}

To exhibit the high light extraction efficiency of our proposed device, we show the calculated far-field emission profile in Fig. 1(b). This emission profile fits to a 2D Gaussian with near-unity overlap ($R^2 = 99.6$), thus demonstrating that the far field emission profile is not altered by the gates. The light extraction efficiency for the device is calculated to be 35\,$\%$, which can be tailored towards unity by optimizing the device architecture including the nanowire shape, removing the dielectric and integrating a gold mirror at the nanowire base\citep{Gregersen:08, Gregersen:16}. 

We now present the underlying theory for calculating the FSS. The breaking of the quantum dot spatial symmetry causes the coupling of electrons and holes via the electron-hole exchange interaction \citep{Bayer2002}. The electron-hole exchange interaction can be decomposed into two parts: short-range (within the wigner-seitz cell) and long-range (outside the wigner-seitz cell)\citep{Bayer2002}. The FSS is mainly determined by the long-range exchange interaction term\citep{PhysRevB.81.245324},
\begin{eqnarray}
\begin{split}
\delta = \int\int d^3r_1 d^3r_2 ([\phi_0^{h}(\vec{r}_2)u_{v\Downarrow}(\vec{r}_2)]^*[\phi_0^{e}(\vec{r_2})u_{c\uparrow}(\vec{r_2})])^*\\ \times\frac{e^2}{4\pi\epsilon\epsilon_0 |\vec{r_1} -\vec{r_2}|}([\phi_0^{h}(\vec{r}_1)u_{v\Uparrow}\phi_0^{e}(\vec{r_1})]^*[ \phi_0^e(\vec{r}_1)u_{c\downarrow}(\vec{r}_1)])
\label{eq:one},
\end{split}
\end{eqnarray}
where $\phi_0^{e}(\vec{r}_i)$, $\phi_0^{h}(\vec{r}_i)$ are the wavefunctions for the lowest electron and hole orbitals as a function of position, $\vec{r}_i$. The Bloch functions of the conduction and valance band are $u_{cs_z}$,$u_{vj_z}$, respectively, with the spin of the electron $(s_z = +1/2$ $ (\uparrow)$, $-1/2$ $(\downarrow))$ and heavy-hole $(j_z = +3/2$ $(\Uparrow)$, $-3/2$ $(\Downarrow))$ resulting in two bright excitonic states of total angular momentum M $=+1(\downarrow,\Uparrow)$ and M $=-1(\uparrow,\Downarrow)$.  To carry out this calculation, we assume a 3D asymmetric parabolic quantum dot potential. In such a model, the ground state electron and heavy-hole wavefunctions can be modeled by a Gaussian,  $\phi_0^{e/h}(\vec{r})=(\frac{1}{\pi^{3/2}l_x^{e/h}l_y^{e/h}l_z^{e/h}})^{1/2} exp\lbrace -\frac{1}{2}[(\frac{x}{l_x^{e/h}})^2 + (\frac{y}{l_y^{e/h}})^2 + (\frac{z}{l_z^{e/h}})^2]\rbrace $.  Equation \ref{eq:one} is solved analytically, giving FSS in terms of the relevant material properties, wavefunctions and quantum dot geometry by the relation\citep{PhysRevB.81.245324},

\begin{eqnarray}
\begin{split}
\delta = K\cdot\beta\cdot\xi(1-\xi)\cdot\frac{\gamma_z}{(l_y^{eh})^3}
\label{eq:FSS3d},
\end{split}
\end{eqnarray}
\noindent where $FSS = 2\vert\delta\vert$; $K = \frac{3\sqrt{\pi}e^2\hbar^2E_p}{(4\pi \epsilon_0)16\sqrt{2}\epsilon m_0 (E_{g}^b)^2}$, is a constant and is dependent on the quantum dot material properties where $E_p$ and $E_{g}^b$ are the conduction-valence band interaction energy and bulk energy gap, respectively, and $m_0$ is the free electron mass; $\beta$ is the e-h wavefunction overlap ($\beta = \vert\langle \Psi_{h} \vert  \Psi_{e}\rangle\vert^2$) where $\Psi_{e}$ and $\Psi_{h}$ are ground state electron and heavy-hole wavefunctions; $\xi = \frac{l_y^{eh}}{l_x^{eh}}$ characterizes the hybridized wavefunction elongation \citep{PhysRevB.76.235304}; and the length parameters $l_{x,y}  ^{eh}$ are the spatial extents of the hybridized e-h wavefunction ($\Psi_{eh} = \Psi_{h}\Psi_{e}$) along the x- and y-axis of the quantum dot; and $\gamma_z$ is the parameter for z confinement. For a quantum dot where the height is much less than its diameter (\textit{i.e.}, $d_z<<d_{x,y}$), the parameter for z confinement ($\gamma_z$) equals one.

This analytical expression implies that there are two main strategies to minimize the FSS. The first approach is to reduce the e-h overlap (\textit{i.e.}, $\beta$) and the second strategy is to make the exciton wavefunction symmetric (\textit{i.e.}, tune $\xi$ to 1). However, reducing $\beta$ will affect the quantum dot brightness since it minimizes the recombination probability of the bright exciton state \citep{Korkusinski2009a}. Hence, the favorable strategy involves the tuning of $\xi$ without compromising the e-h overlap, $\beta$.

We calculate the FSS for a GaAs quantum dot with dot diameter of 30\,nm and dot height of 3\,nm. The dot is embedded in an $Al_{0.33}Ga_{0.67}As$ nanowire shell (thickness of 110\,nm), which is surrounded by an $ Al_2O_3$ dielectric layer (thickness of 150\,nm). Finally, four gold electrical contacts (width of 200\,nm) are defined in the plane of the quantum dot to apply a quadrupole electric potential. Refer to Fig.1(a) for a schematic view of the proposed device architecture. 

\begin{figure}[t!]
\includegraphics{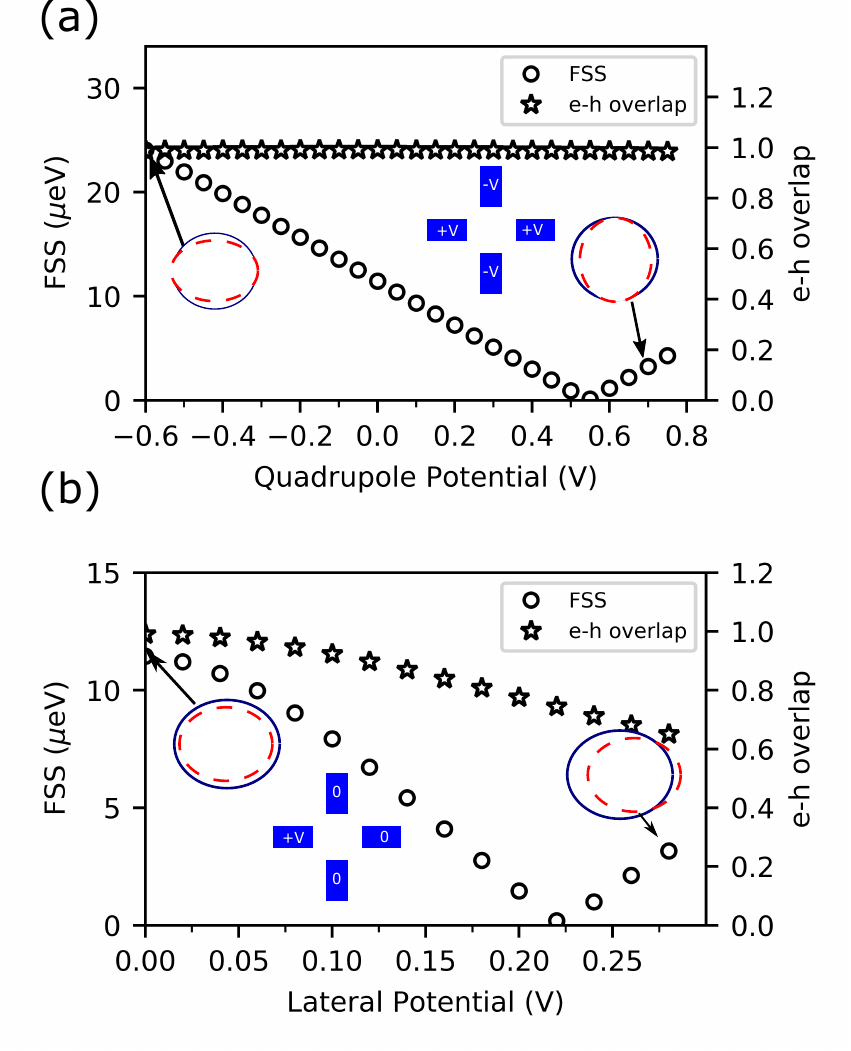}
\caption{\label{fig:quadFSS}(a) Calculated FSS (circles, left axis) and e-h overlap (stars, right axis) as a function of quadrupole potential. (b) Calculated FSS (circles, left axis) and e-h overlap (stars, right axis) as a function of lateral potential. Insets: Schematic view of applied quadrupole (in (a)) and lateral potential (in (b)). The probability density of the electron (solid blue) and hole (dashed red) wavefunctions are schematically shown by the ellipses.}
\end{figure}

To demonstrate that the proposed device allows for FSS correction without compromising the quantum dot brightness, we have performed a numerical simulation using nextnano \citep{nextnano} to solve the two-dimensional Schr\"{o}dinger-Poisson equation self-consistently using an effective mass approximation. The material parameters used for the GaAs quantum dot is listed in Table 1. In our calculations we assume the ground state is pure heavy-hole. This assumption is in-line with previous theoretical and experimental results\citep{Bester2003,Gerardot2008} where the ground state is dominantly heavy-hole. Ignoring the third dimension is justified for strong confinement $(\gamma_z =1)$, where the dot height is much less than its diameter (\textit{i.e.}, $d_z<<d_{x,y}$). In our simple model, we only consider the geometric quantum dot asymmetry as this is typically the dominant source of FSS \citep{Abbarchi2008}. Thus, we have modified the quantum dot shape from a circle to an ellipse with 7$\%$ elongation along the major axis (\textit{i.e.}, $L_x/L_y = 1.07$). The ground state electron and heavy-hole wavefunctions are then computed from the solution of the two-dimensional Schr\"{o}dinger-Poisson equation from which we calculate the hybridized e-h wavefunction. Using equation \ref{eq:FSS3d}, we then calculate the FSS. For this quantum dot elongation of 7$\%$, the calculated FSS is $11$ $\mu$eV, which is typical for quantum dots. 

\begin{table}
\caption{\label{tab:table1} Material parameters for GaAs quantum dot  }
\begin{ruledtabular}
\begin{tabular}{ccccc}
 $E_{g}^b$ (eV) & $E_p$ (eV) & $m^*_e$ ($m_o$) & $m^*_{hh}$ ($m_o$)& $\epsilon$  \\
\hline
 1.519 & 23 & 0.067 & 0.5 & 12.5
\end{tabular}
\end{ruledtabular}
\end{table}

Fig. 2(a) and (e) show a schematic view of the applied quadrupole electric potential for two different configurations. Configuration 2(a) corresponds to a positive potential applied to the top and bottom gates, and a negative potential with the same magnitude applied to the left and right gates, while it's the opposite for configuration 2(e). Here, the solid blue and dashed red ellipses represent the single particle wavefunctions of the electron and hole, respectively, under the applied electric potential. The potential profile is plotted for these two configurations from the solution of the Schr\"{o}dinger-Poisson equation in Fig. 2(b) and (f). The black circles represent the $GaAs$ quantum dot (smallest circle at center), $Al_{0.33}Ga_{0.67}As$ nanowire (middle circle), and the $Al_2O_3$ dielectric shell (largest circle). The contour plot of the electric potential shows that it is zero inside of the quantum dot, which is essential for maintaining strong e-h overlap of the bright exciton state. 
\begin{figure*}[t!]
\centering
\includegraphics[scale = 1]{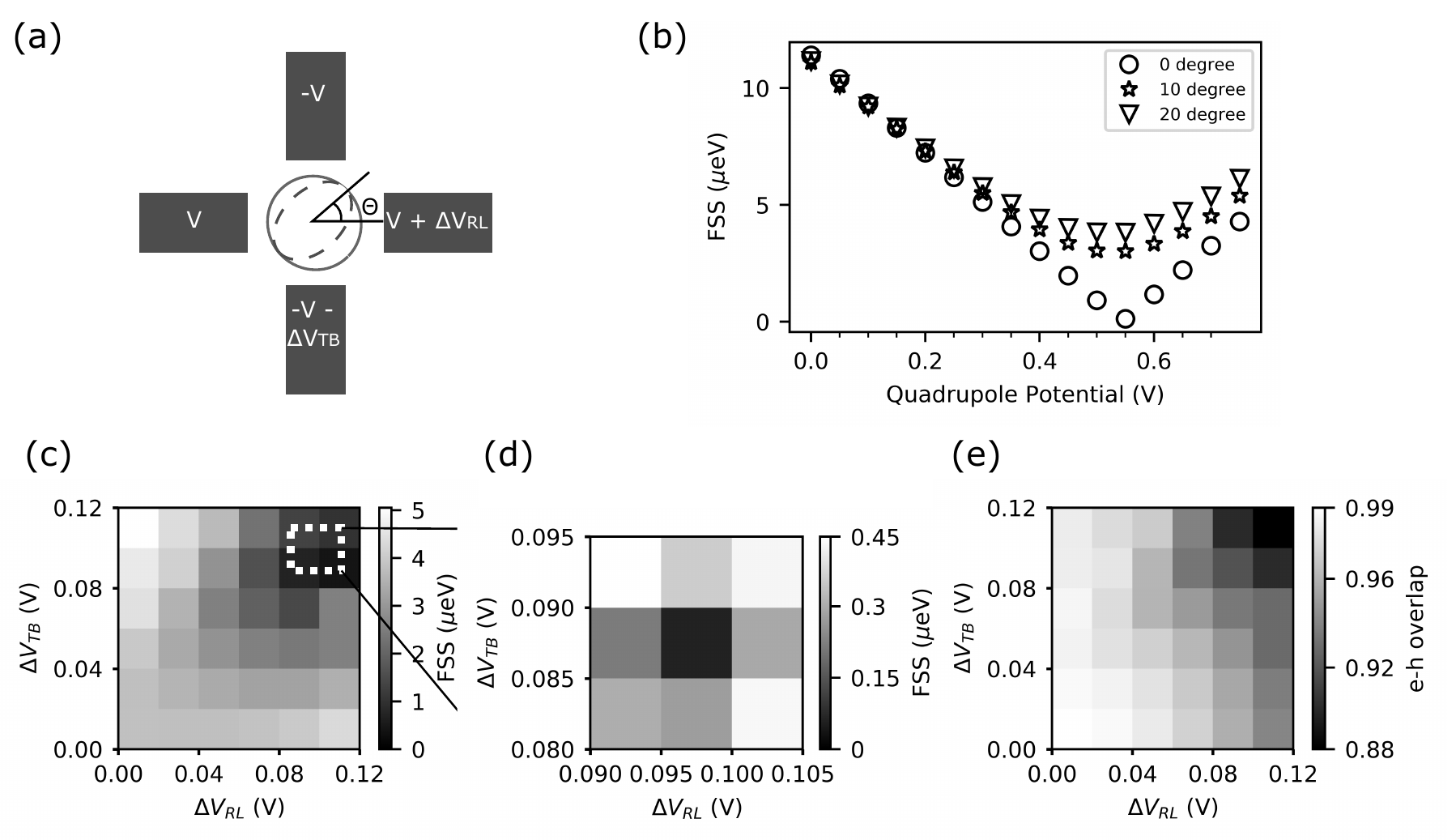}
\caption{\label{fig:degree}Universal FSS tuning. (a) Schematic view of an applied asymmetric quadrupole potential. Inset: electron (solid grey) and hole (dash grey) wavefunctions of a quantum dot with asymmetric axis aligned at an angle $\theta$ with respect to the gates along the x axis. (b) Calculated FSS as a function of quadrupole potential with the quantum dot major axis oriented at three different angles with respect to the gates along the x-axis ($\theta = 0^\circ, 10^\circ,$ and $20^\circ$).   (c) Greyscale plot of FSS as a function of $\Delta V_{RL}$ and $\Delta V_{TB}$ for $\theta = 20^0$. (d) High resolution greyscale plot of the zoomed-in region (white dotted line box from (c)) showing near-zero FSS. (e) Greyscale plot of e-h overlap as a function of  $\Delta V_{RL}$ and $\Delta V_{TB}$ for $\theta = 20^0$. }
\end{figure*}

Fig. 2(c) and (d) show the 2D probability densities of the ground state electron ($\vert \Psi_{e} \vert ^2$) and heavy-hole ($\vert \Psi_{h} \vert ^2$), respectively, for the applied potential configuration from Fig. 2(a). These plots have been fit with 2D Gaussian functions to extract the standard deviations along their major ($\sigma_x$) and minor ($\sigma_y$) axes, and to obtain the single particle wavefunction elongation factor, $\epsilon = \sigma_x/\sigma_y $. We find that the hole wavefunction is stretched along the x-axis ($\epsilon$ = 1.17) and is perturbed much more than the electron wavefunction which is slightly stretched along the y-axis ($\epsilon$ = 0.96). In contrast, Fig. 2(g) and (h) show the 2D probability densities of the electron and hole, respectively, for the applied potential configuration from Fig. 2(e). These plots show an opposite trend to those in Fig. 2(c) and (d): $\vert \Psi_{h} \vert ^2$ is now spread along the y-axis ($\epsilon = 0.90$) instead of the x-axis and $\vert \Psi_{e} \vert ^2$ is spread along the x-axis ($\epsilon = 1.06$) instead of the y-axis. The higher perturbation of the heavy-hole wavefunction is expected since heavy-holes are much more localized than electrons due to their larger effective mass.

In Fig. 3(a) we plot the calculated FSS (circles, left axis) as a function of the quadrupole potential (V), whereby a negative quadrupole potential of -0.5\,V represents the configuration from Fig. 2(a) and a positive quadrupole potential of +0.7\,V represents the  configuration from Fig. 2(e). The trend demonstrates that the FSS can be tuned to zero by applying a quadrupole potential. The calculated e-h overlap for the quadrupole field is also shown in Fig. 3(a), represented by stars (right axis). Remarkably, the quadrupole field maintains near-unity e-h overlap, with a value above 99\,$\%$ over the entire range of the applied field, thus conserving the quantum dot brightness. This desirable feature is in stark contrast to previous work where an applied lateral field was used and the e-h overlap had to be drastically reduced to remove the FSS
\citep{Gerardot2007a,Korkusinski2009a,Kowalik2005,
Kowalik2007}.

To directly compare the two approaches (quadrupole and lateral electric field), we modeled the device under an applied lateral field with a positive potential on the left gate while keeping all other gates at zero potential (Fig. 3(b)). In this case, the FSS (circles, left axis) can be corrected for by applying V = 0.225\,V, but the e-h overlap (stars, right axis) is reduced to 74\,$\%$ at this potential (Fig. 3(b)), thus compromising the quantum dot emission brightness. We re-emphasize that in the case of the lateral electric field the gates must be aligned along the quantum dot dipole orientation, otherwise the FSS cannot be removed. 

The device architecture that we have modeled represents an ideal case where the four electrical gates are perfectly aligned along the quantum dot dipole orientation. In practical devices; however, the quantum dot asymmetric axis is randomly oriented from dot to dot. Moreover, there is also the challenge of fabricating the device with the required precision. In such cases, the quantum dot asymmetry will be misaligned with respect to the electrical gates. To model this misalignment, we have simulated two additional dipole orientations with the quantum dot major axis oriented at an angle of $\theta = 10^\circ$ and $\theta = 20^\circ$ with respect to the gates along the x-axis (Fig. 4(a)). Fig. 4(b) shows the calculated FSS as a function of quadrupole potential for $\theta = 0^\circ, 10^\circ$ and $20^\circ$. The minimum FSS is obtained for an applied quadrupole potential of V = 0.5 V (for $\theta = 10^{0},20^0$). Clearly, there exists a non-zero minimum bound to the FSS for $\theta = 10^\circ$ and $\theta = 20^\circ$, whereas the FSS can reach zero for $\theta = 0^\circ$ 
%(perfect alignment of the quantum dot dipole orientation to the gates). 
 
To reduce the minimum bound of the FSS further when the quantum dot dipole orientation is not aligned with the electrical gates ($\theta = 20^0$), we modify the applied quadrupole potential by increasing the potential on the right gate by $\bigtriangleup V_{RL}$ and decreasing the potential on the bottom gate by $\bigtriangleup V_{TB}$. The calculated FSS is plotted as a function of $\bigtriangleup V_{RL}$ and $\bigtriangleup V_{TB}$ in Fig. 4(c). In Fig. 4(d) we zoom into the region of the white dotted box of Fig. 4(c) and find near-zero FSS (0.05 $\mu$eV) at $\bigtriangleup V_{RL}$ = 0.095\,V and $\bigtriangleup V_{TB}$ = 0.085\,V.  

The e-h overlap is also plotted as a function of $\bigtriangleup V_{RL}$ and $\bigtriangleup V_{TB}$ in Fig. 4(e). An overlap of 90$\%$ is obtained for the modified quadrupole potential near-zero FSS. This high e-h overlap with close to zero FSS is desired for a high-efficiency entangled photon source with near-unity fidelity. 

%Our approach to erase the FSS with a quadrupole field is also promising for building a quantum repeater \citep{Duan2001}. The key ingredients for building a quantum repeater is a tunable source of entangled photons with near-unity efficiency, fidelity and  indistinguishability. We envision that the quantum dot emission wavelength can be tuned for entanglement swapping by adding an additional back-gate. This same back gate can also be used to control the electrostatic environment of the quantum dot for achieving near-unity  indistinguishability\citep{Somaschi2015}.  

In summary, we showed that an applied quadrupole field to a single quantum dot in a photonic nanowire results in zero FSS without compromising the quantum dot brightness.  With our approach we envision high-efficiency entangled photon sources with near-unity fidelity are within reach. Such sources are the missing building block of quantum networks and can also be used as a new source in quantum communication, quantum sensing, quantum optics and integrated quantum optics on a chip.

%We would like to thank Stefan Birner for the help with nextnano and Andreas Fognini for scientific discussions. This research was undertaken thanks in part to funding from the Canada First Research Excellence Fund, NSERC and Industry Canada. 

\bibliography{mybib}
\bibliographystyle{aipnum4-1}

\end{document}